# A Model-Independent Determination of Red Noise in Pulsar Timing Arrivals

**Reinabelle Reyes[1] and Christopher C. Bernido[2*]**

[1]National Institute of Physics, University of the Philippines, Diliman, Quezon City 1101, Philippines

[2]Research Center for Theoretical Physics, Central Visayan Institute Foundation, Jagna, Bohol 6308, Philippines

**\* Correspondence:**
C. C. Bernido
cbernido.cvif@gmail.com

**Abstract**

Noise is ubiquitous in pulsar signals where red noise has been attributed to effects arising from the interstellar medium, spin noise, and pulsar mode changes, among others. The red noise, however, has not been detected in all pulsars. Using the dataset from the North American Nanohertz Observatory for Gravitational Waves (NANOGrav), we investigate twenty-three pulsars and show that an evaluation of the mean square deviation and probability distribution of timing residuals can provide a straightforward way of determining the presence of red noise. The results agree with those reported in the literature. The model-free method presented could complement the normally more sophisticated model-dependent way of determining red noise in timing residuals.

## 1 Introduction

An intensified investigation of noise in pulsar timing arrivals has been driven by efforts to detect gravitational waves (Antoniadis et al 2022; Goncharov et al 2021; Abbott et al 2016; Arzoumanian et al 2018; Lam et al 2016; Dahal 2020; Hazboun et al 2019; Lommen 2015; Wang 2015; Sarah Burke-Spolaor, et al., 2019; Woods 2022). The Gaussian white noise is known to dominate at short time scales and high fluctuation frequency. On the other hand, at longer time scales and lower fluctuation frequency, red noise is detected in pulsar timing observations (Goncharov 2021; Shannon and Cordes 2010). Red noise from pulsars could confound the detection of gravitational waves which may also manifest as a red noise background (Hazboun et al 2020; Chalumeau et al 2022). In a recent study of the 11-year NANOGrav dataset for 45 millisecond pulsars, only 11 pulsars showed significant red noise using timing-model fits (Arzoumanian et al 2018). A Bayes-factor analysis was used to find significant evidence of red noise in these 11 pulsars. We shall use the results of this study to check the outcome of the method presented in this paper.

We obtain the timing residuals for 23 pulsars from the 11-year data of NANOGrav. The mean square displacement (MSD) of timing residuals is evaluated for each pulsar from 2008 to 2016 and classified according to their behavior. The corresponding probability displacement distribution for each pulsar is likewise evaluated and classified according to significant characteristics revealed at different time scales. The combined classification of the MSD and probability displacement distribution for the

23 pulsars unveils an underlying characteristic for pulsars with red noise. Despite the errors in pulsar timing measurements, a compelling pattern is revealed showing which pulsars would exhibit a red noise. The results agree with those reported in a previous work (Arzoumanian et al 2018). This model-independent determination of red noise in pulsar signals could be a useful addition to the existing arsenal already employed in the investigation and understanding of noise in pulsar timing observations. Section 2 presents the data and method used. We discuss the results in Section 3.

## 2 Data and Methods

### 2.1 Timing Residuals

The timing residuals for 23 pulsars with 820 MHz band data were obtained from the 11-year NANOGrav data (Arzoumanian et al 2018). Although the raw data for 820 MHz band goes back to 2005, the interval between data points for the earlier data has larger gaps. We, therefore, restrict our investigation to data from 2008 onwards. Those with complete data cover eight years from 2008 to 2016. To get a time series with uniform time intervals set to 30 days, an interpolation was applied. The time point being interpolated is set to the data point nearest to it. The timing residuals in microseconds for the 23 pulsars are shown in Fig. 1.

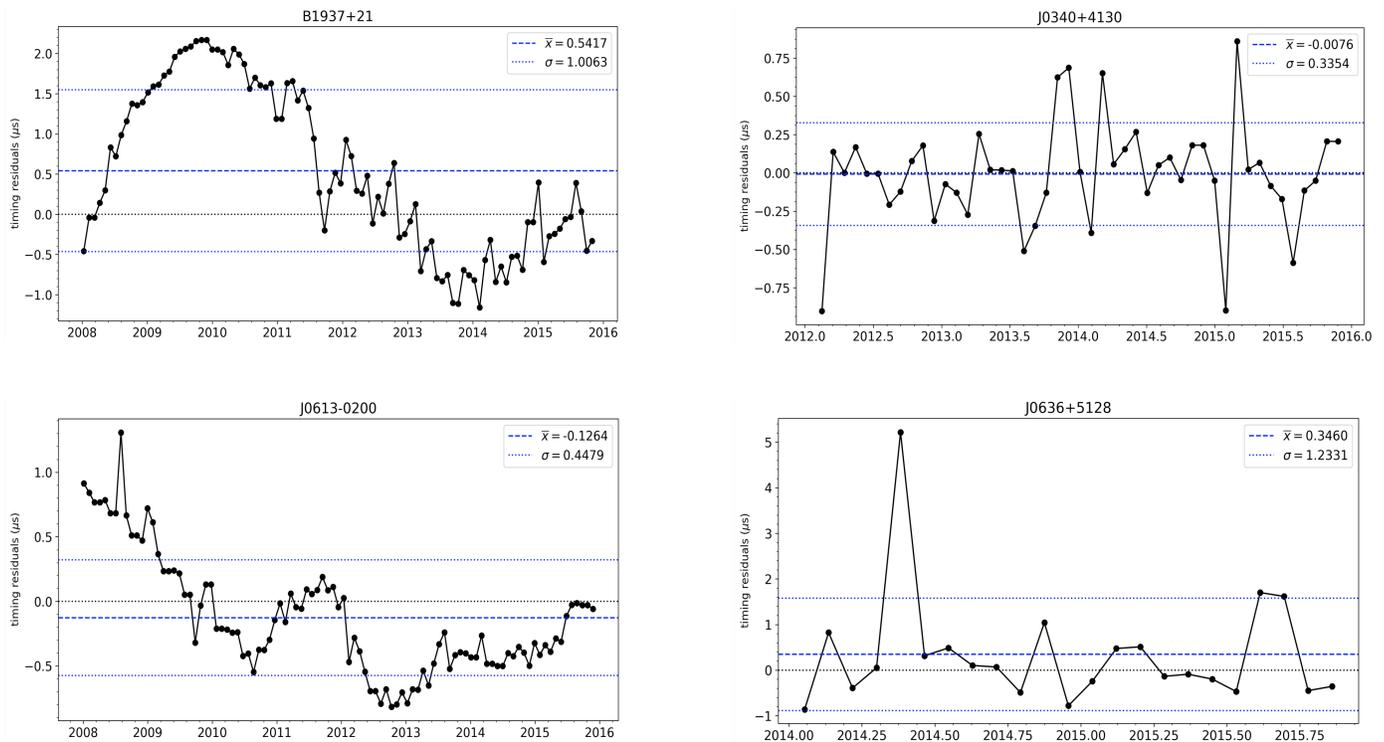

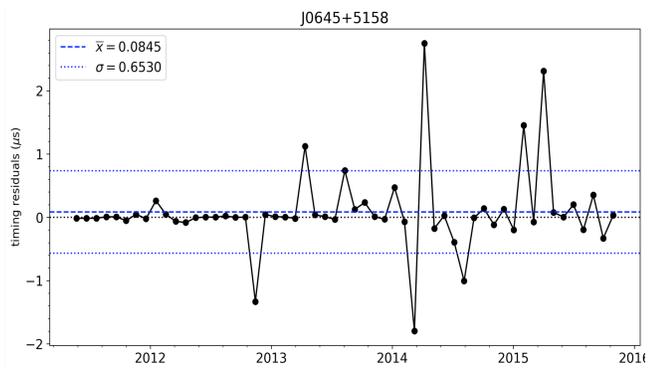
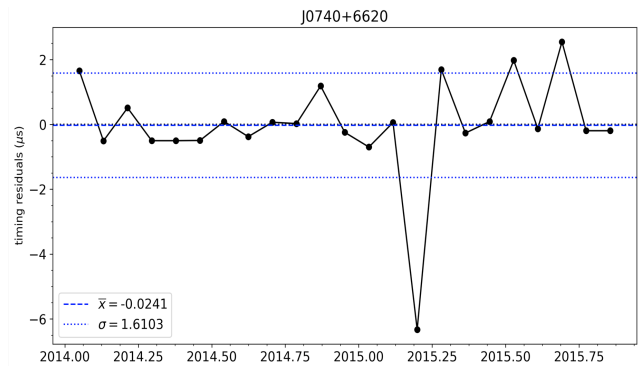
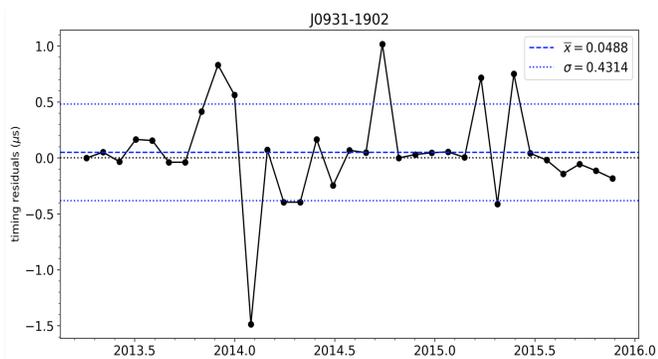
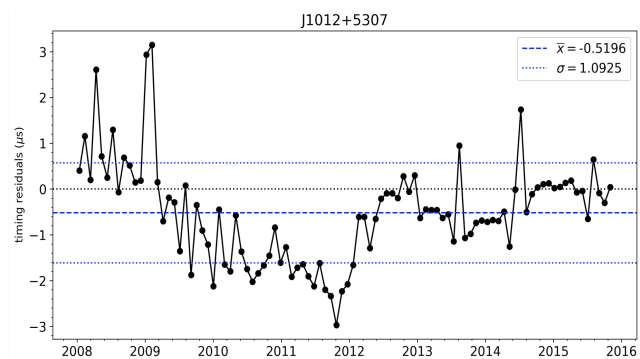
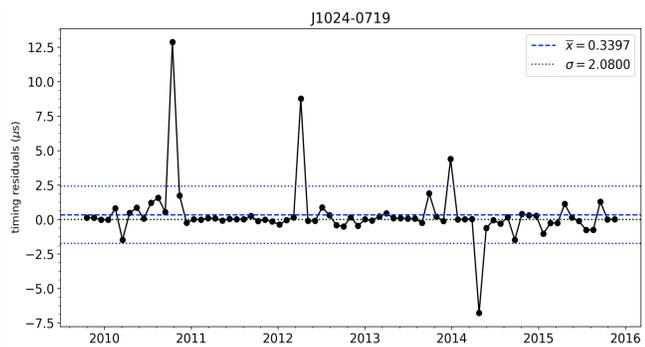
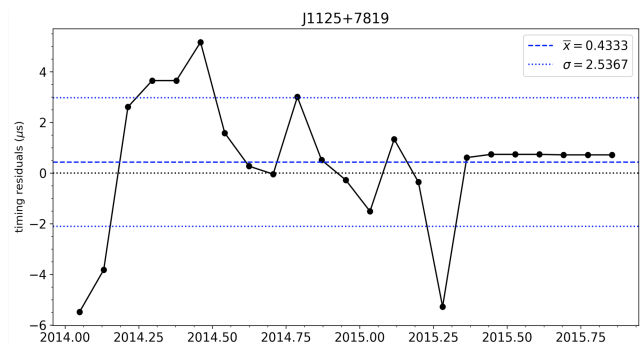
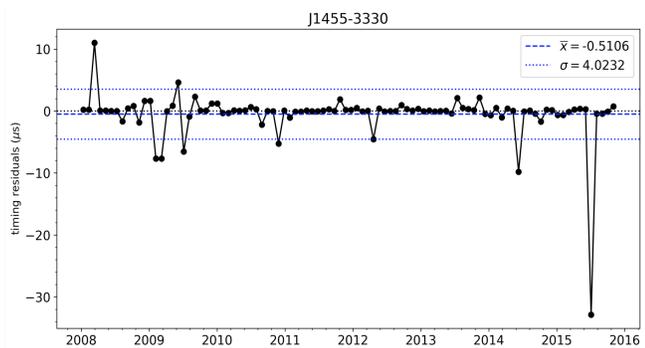
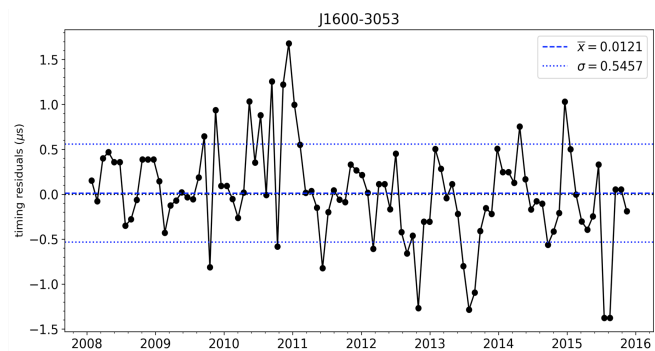

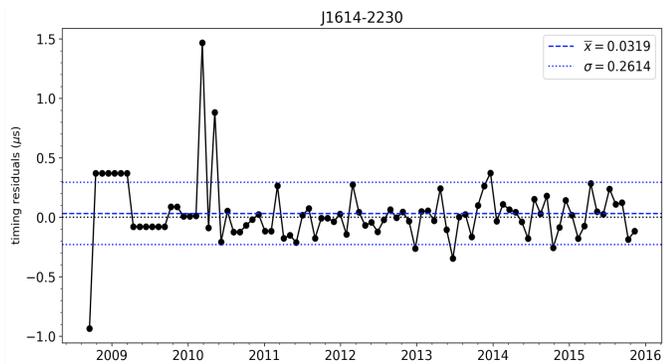
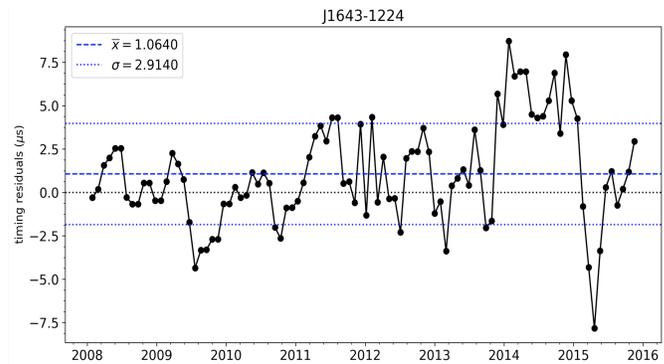
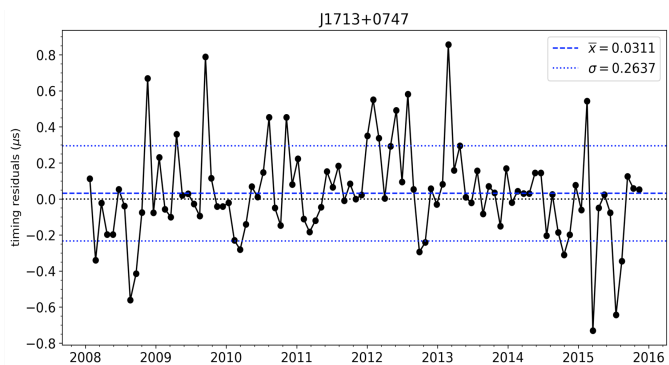
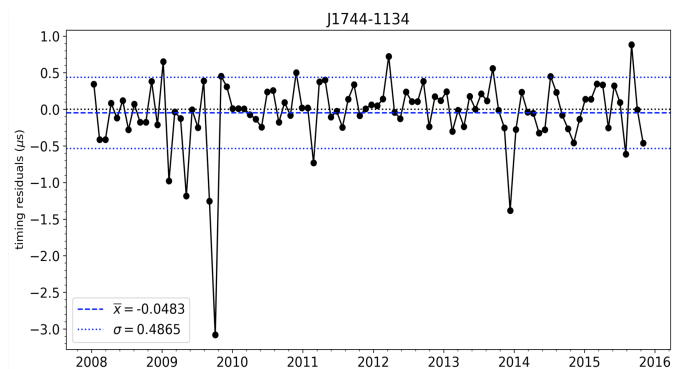
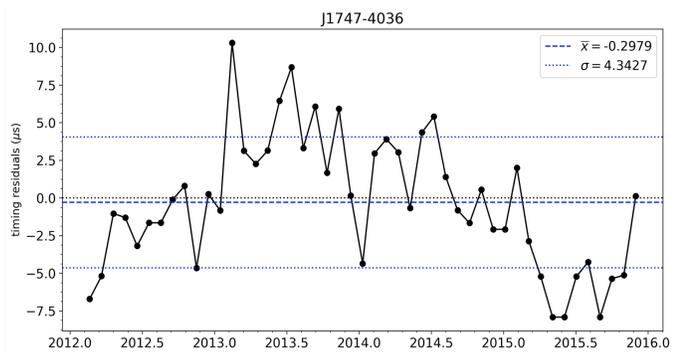
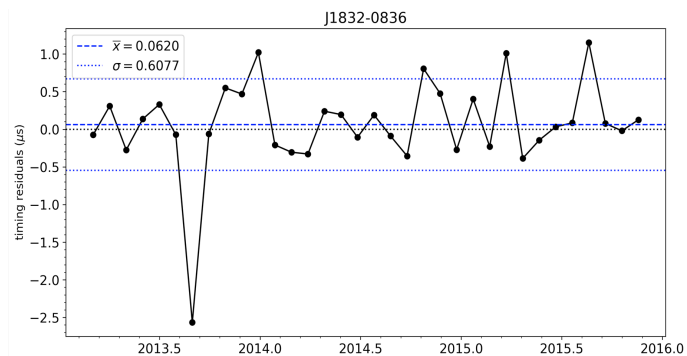
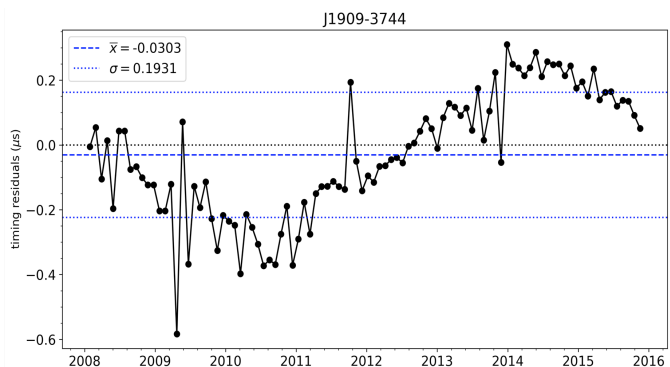
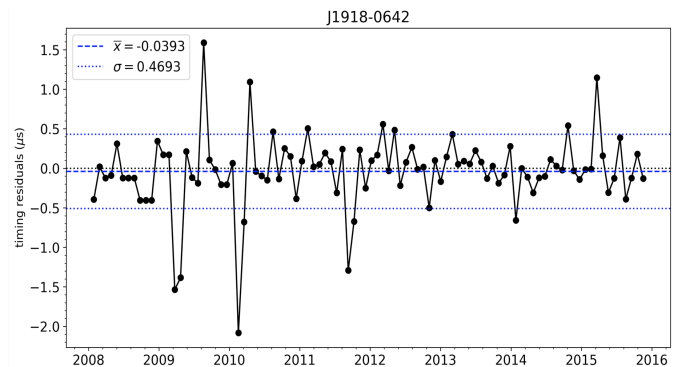

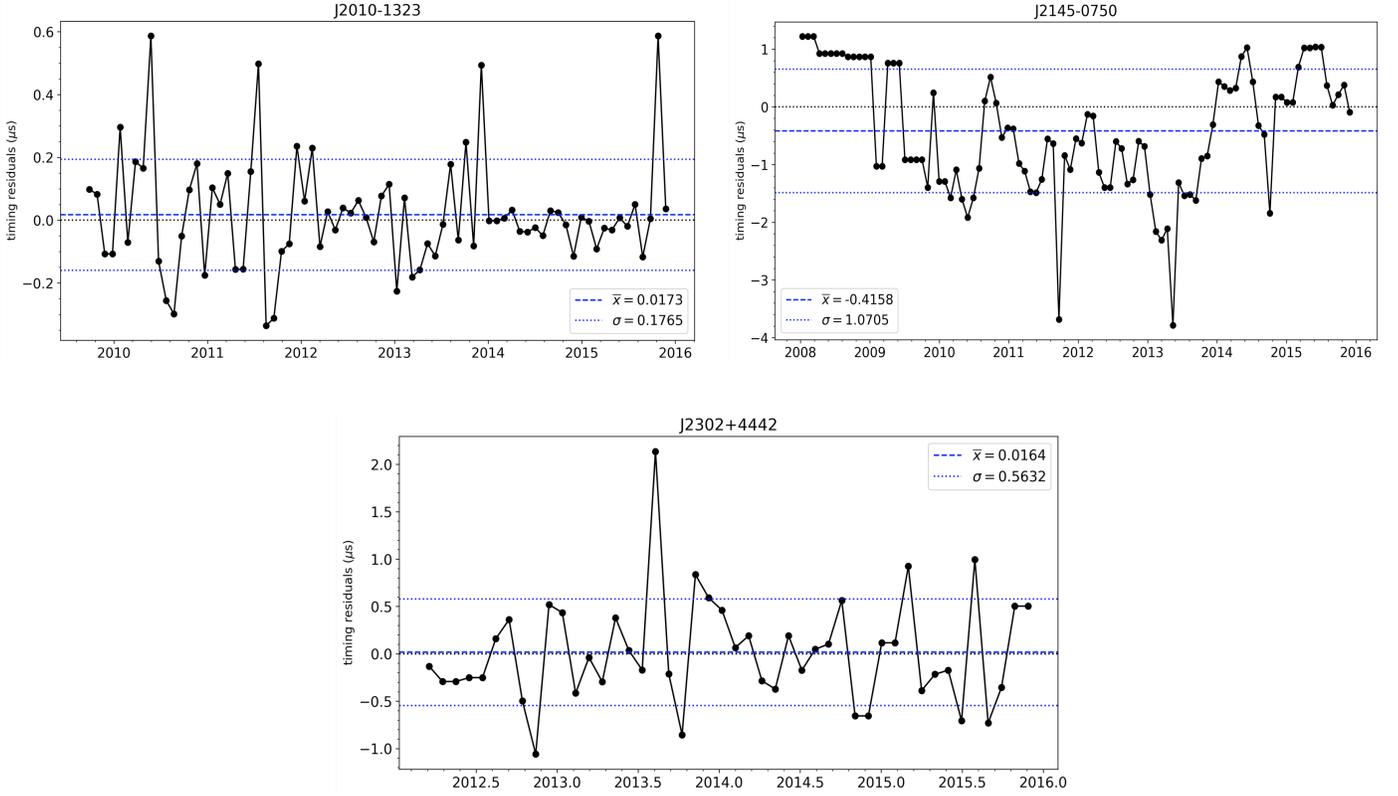

**Figure 1:** Timing residuals for 23 pulsars in the 820 MHz band. The name of the pulsar is shown at the top of each plot. Dashed blue line show the mean value and dotted blue lines show +/- 1 standard deviation from the mean.

### 2.2 Mean Square Displacements

We shall designate $N$ as the total number of datapoints in the timing residuals, and $x_j$ as the value of the datapoint at time $t_j$ with, $j = 0, 1, 2, ..., N$. Hence, the value of the timing residual at the initial time $t_0$ would be, $x_0 = x(t_0)$, and the last value at time $t_N$ is given by $x_N = x(t_N)$. The mean square displacement for timing residuals is evaluated using,

$$\text{MSD}(\Delta) = \frac{1}{N-\Delta+1} \sum_{j=0}^{N-\Delta} (x_{j+\Delta} - x_j)^2 \ , \qquad (1)$$

where, $\Delta < N$, is the lag time. The lag time $\Delta$ is the interval or separation between any two data points. By evaluating Eq. (1) for each lag time $\Delta$, a plot of MSD versus lag time can be generated. Larger values of lag times, however, result in fewer deviations squared that are summed in Eq. (1). To obtain statistically significant plots, a cut-off value of $\Delta$ is imposed. The plotted MSD is up to $\sim N/2$ and are shown in Figs. 2 and 3.

### 2.3 Probability Displacement Distribution

From the dataset, we again consider $x_j$ which is the value of a timing residual at time $t_j$. We look at the difference, or displacement, $x_{j+\Delta} - x_j$, for a fixed lag time $\Delta$, and count the number of similar displacements as we go through different values of $j$. We can then plot a probability displacement distribution showing the frequency of appearance of similar values for a given Δ. Figures 4 to 6 show the probability displacement distribution for each pulsar as lag time Δ is increased.

## 3 Results and Discussion

### 3.1 Two Types of MSD

Two different types of MSD emerge for pulsars from the timing residuals shown in Section 2. One class of MSD increases with lag time, and the other type remains essentially flat as lag time increases. We label the MSD's accordingly as the Increasing MSD shown in Fig. 2, and the Flat MSD in Fig. 3.

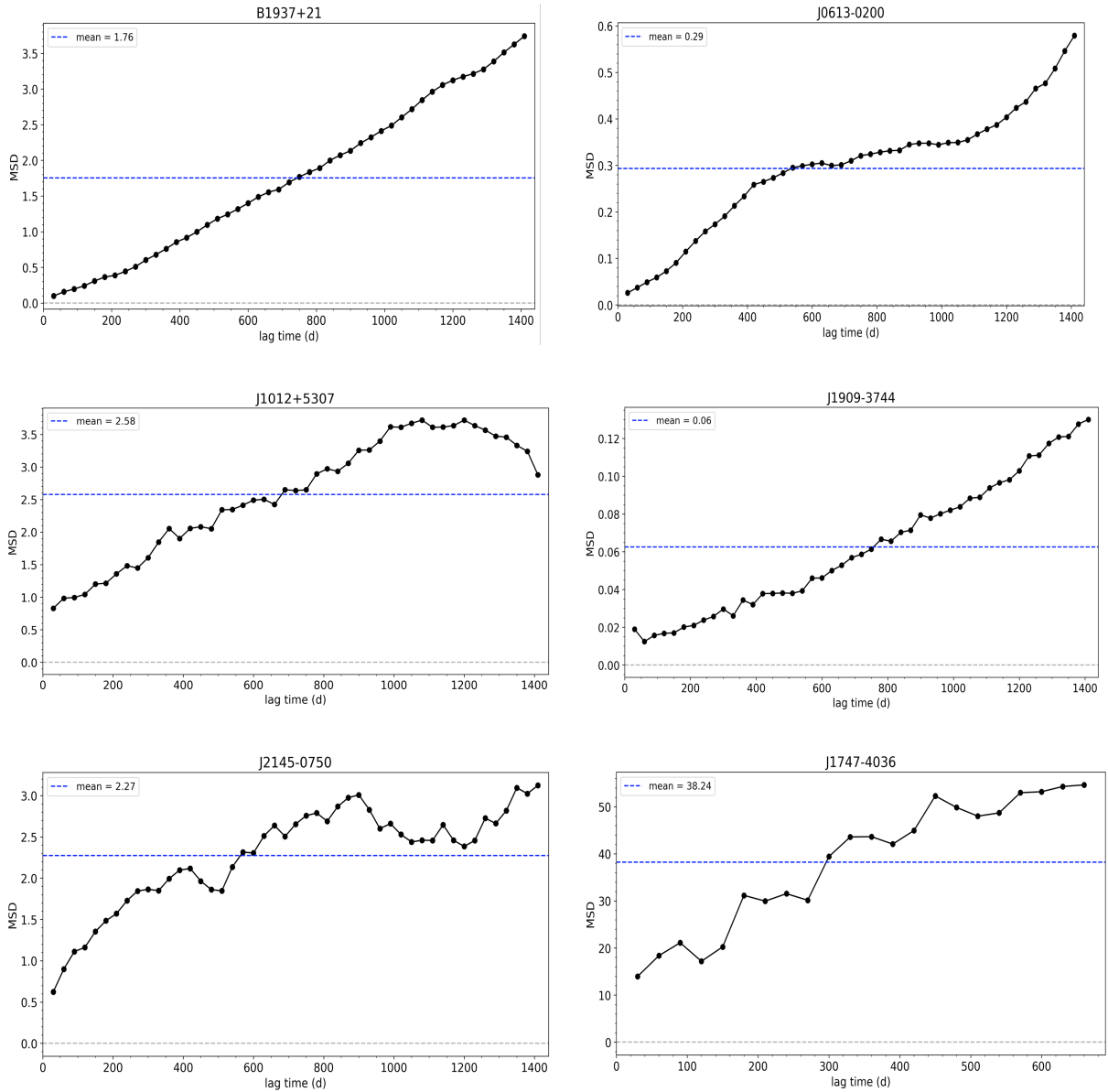

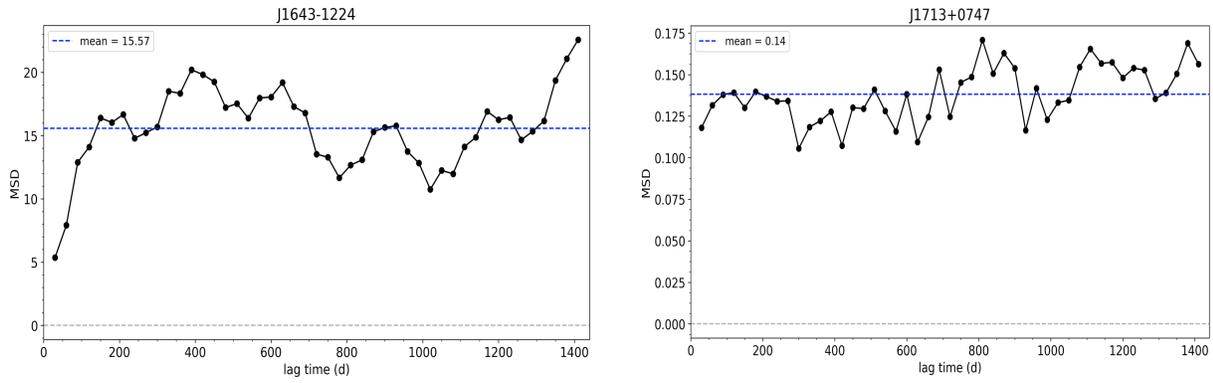

**Figure 2:** Increasing MSD for pulsars B1937+21, J0613-0200, J1012+5307, J1909-3744, J2145-0750, J1643-1224, and J1713+0747. Pulsar J1713+0747 (bottom right) has the smallest amplitude and lowest Bayes factor making it difficult to detect red noise (Arzoumanian et al 2018). All eight pulsars exhibit red noise.

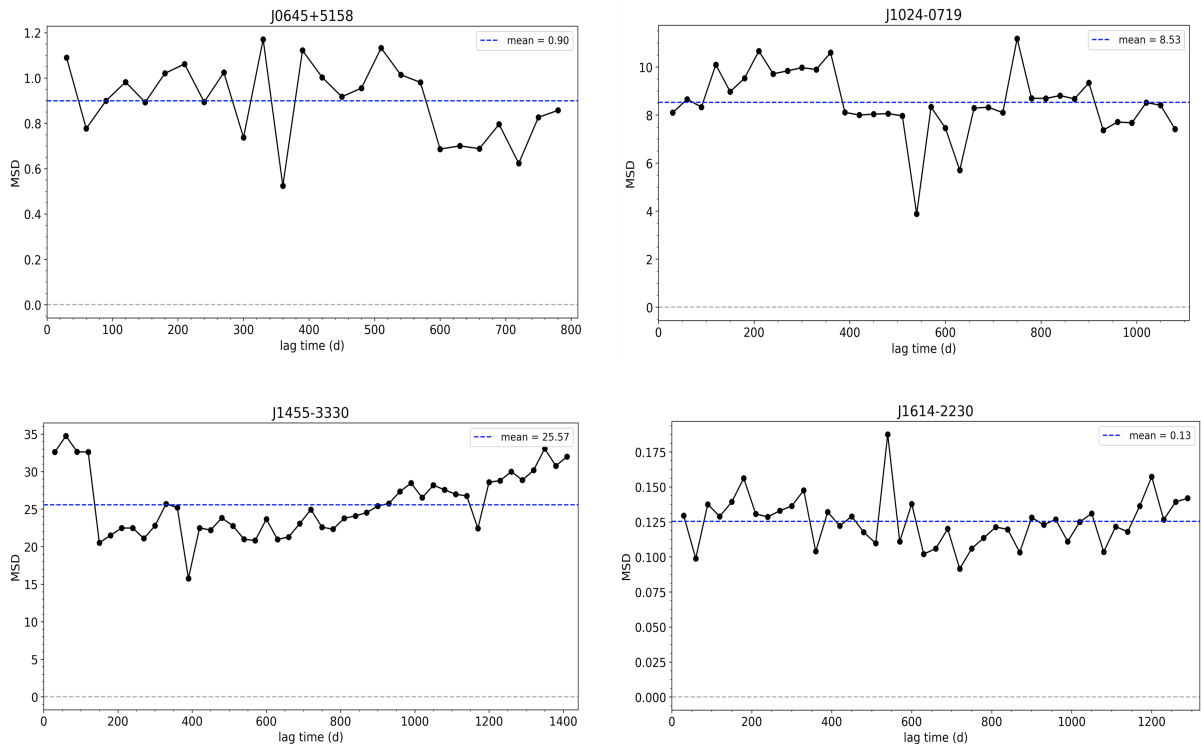

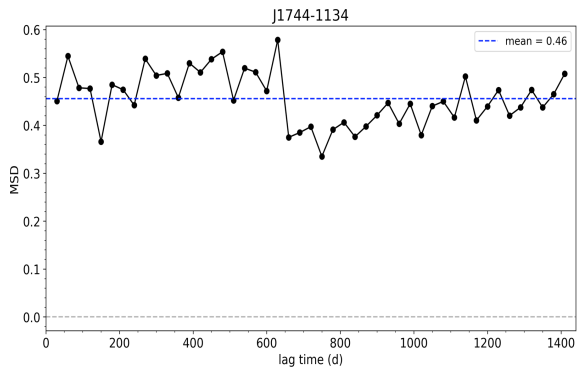
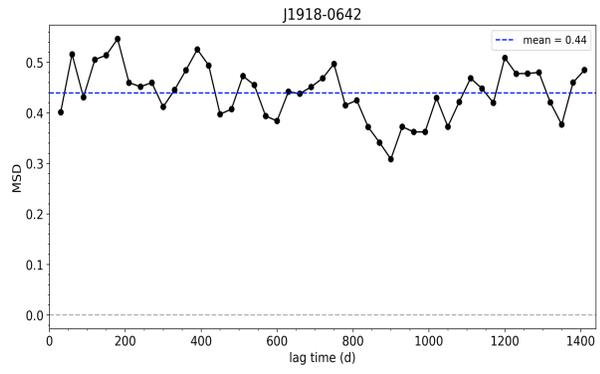
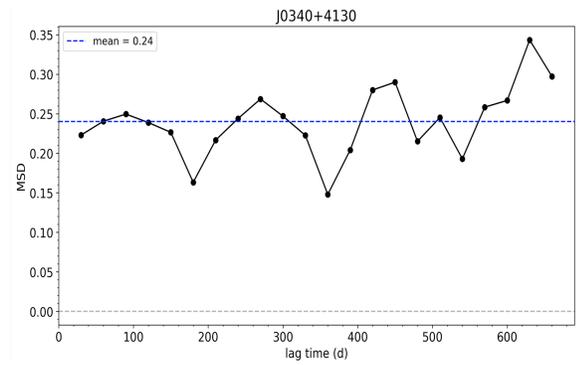
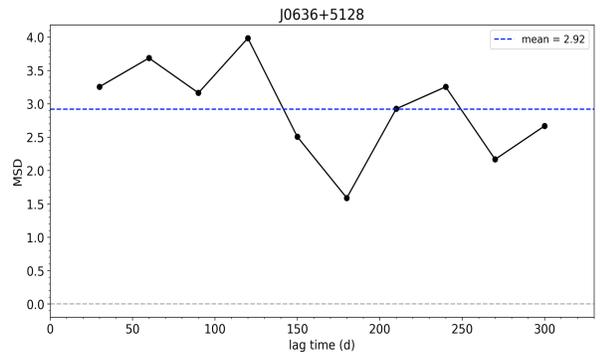
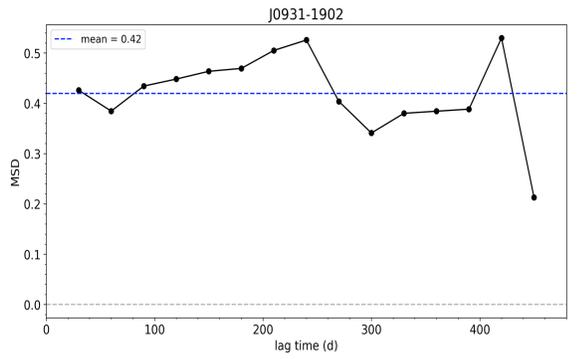
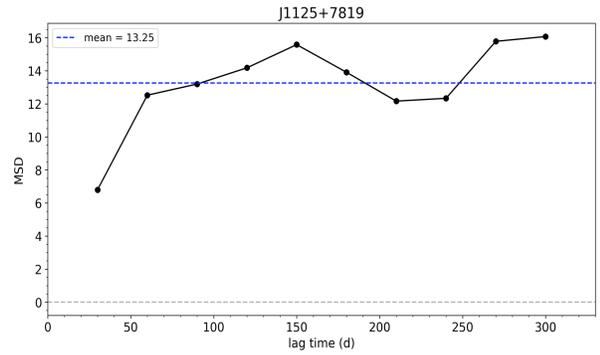
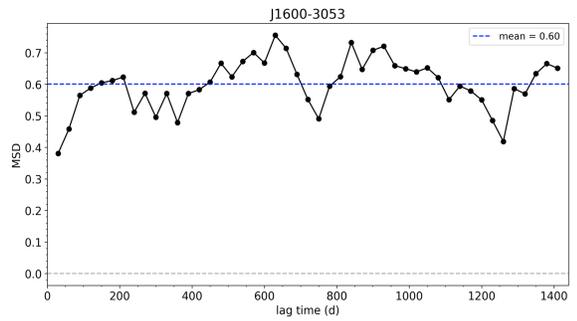
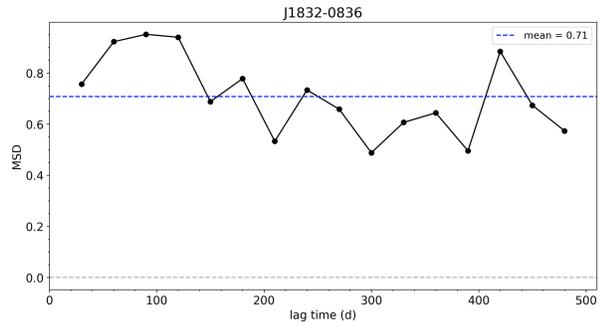

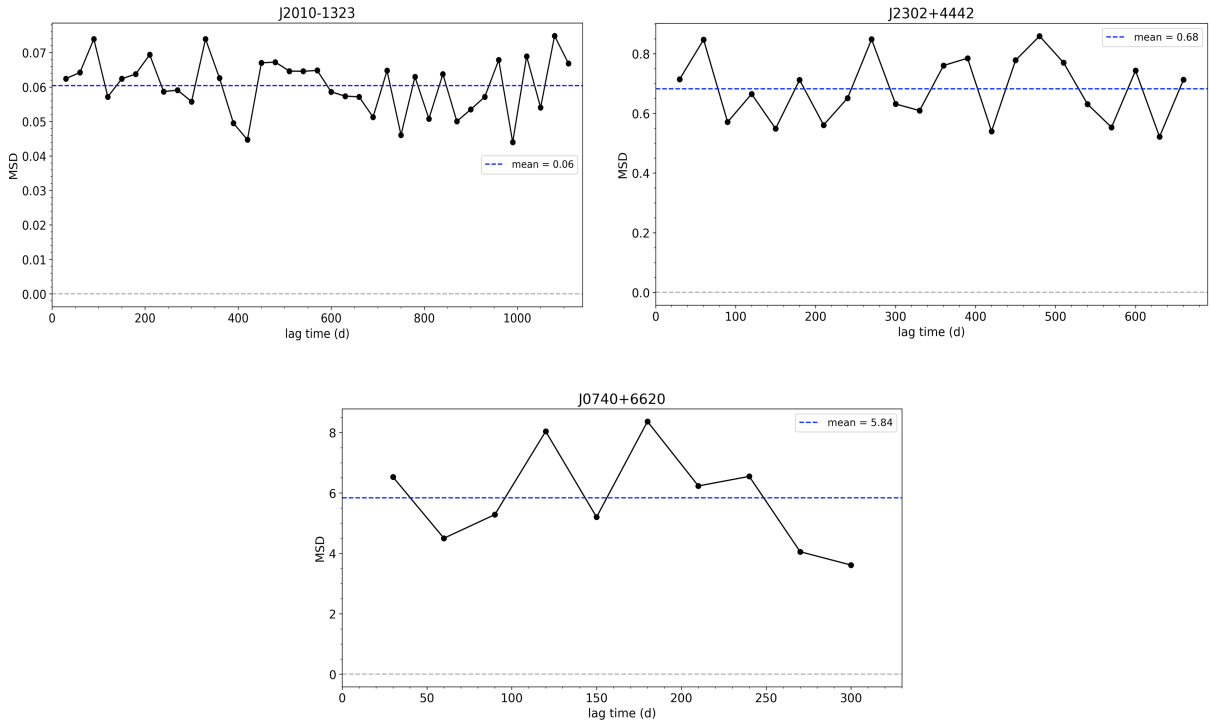

**Figure 3:** Flat MSD for pulsars J0645+5158, J1024-0719, J1455-3330, J1614-2230, J1744-1134, J1918-0642, J0340+4130, J0636+5128, J0931-1902, J1125+7819, J1600-3053, J1832-0836, J2010-1323, J2302+4442, and J0740+6620. All fifteen pulsars do not exhibit red noise.

## 3.2 Three Types of Probability Displacement Distributions

The probability displacement distributions for pulsars reveal three types of profiles as the lag time increases. We label as Type 1 those pulsars with timing residuals that exhibit a probability displacement distribution with a single peak despite the increasing values of lag times (see, Fig. 4). Type 2 pulsars would have timing residuals characterized by a probability distribution that has a single peak only at a short lag time, and develops multiple peaks at longer lag times (see, Fig. 5). Type 3, on the other hand, shows multiple peaks and a broad distribution regardless of the value of the lag time (see, Fig. 6).

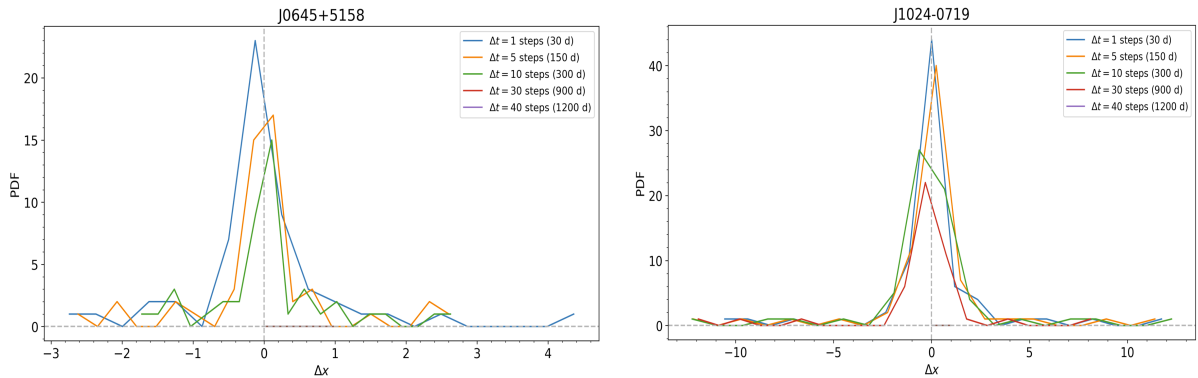

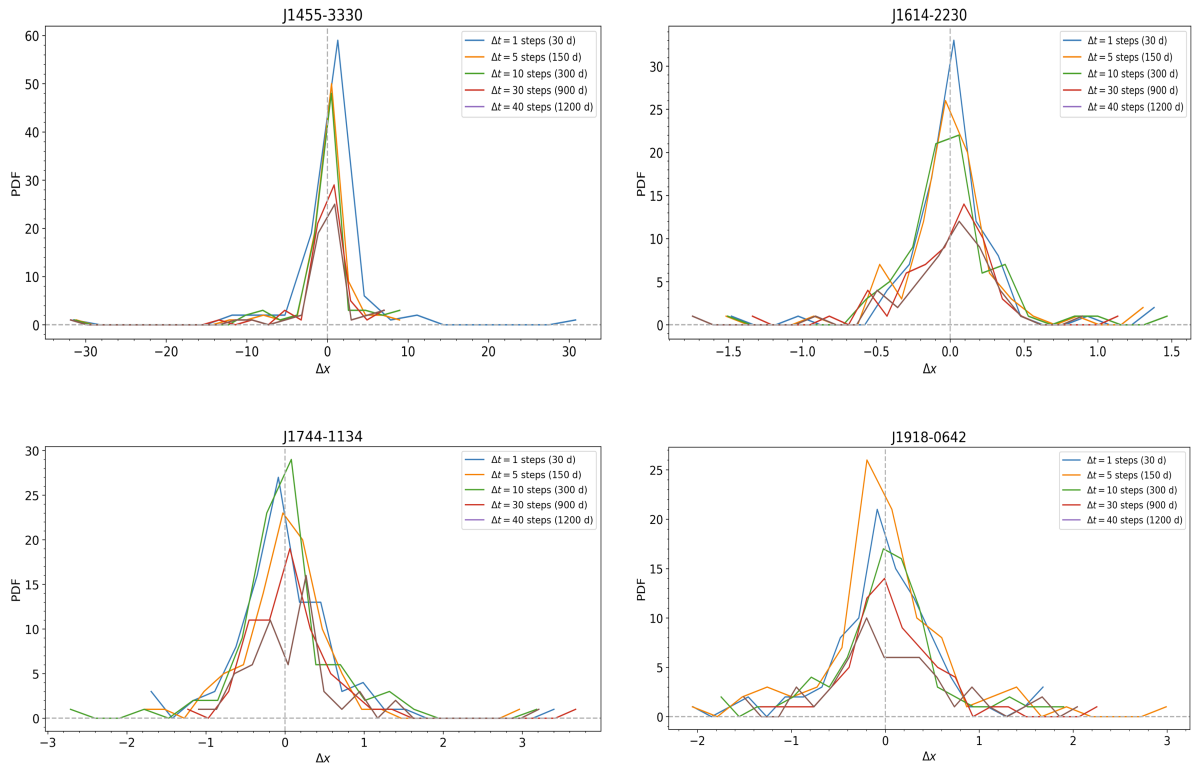

**Figure 4:** Type 1 probability distributions are single-peaked regardless of the value of the lag time. Pulsars of ths type are: J0645+5158, J1024-0719, J1455-3330, J1614-2230, J1744-1134, and J1918-0642. All six pulsars do not have red noise.

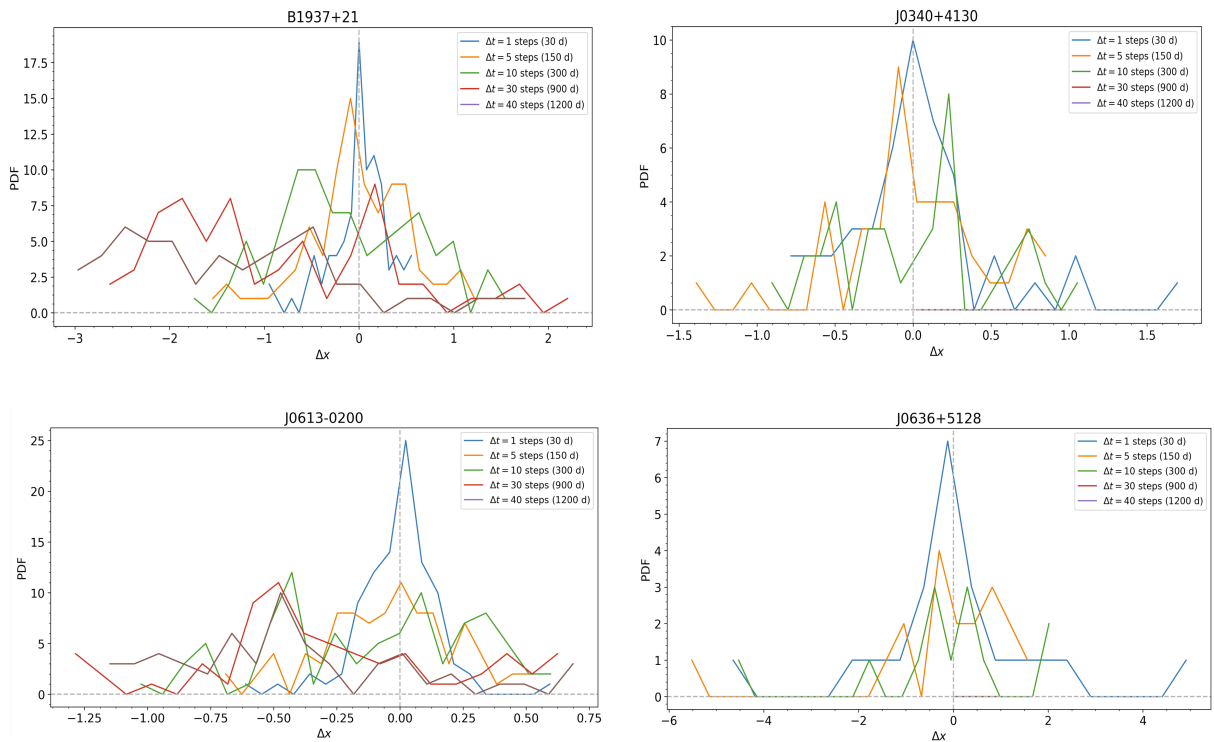

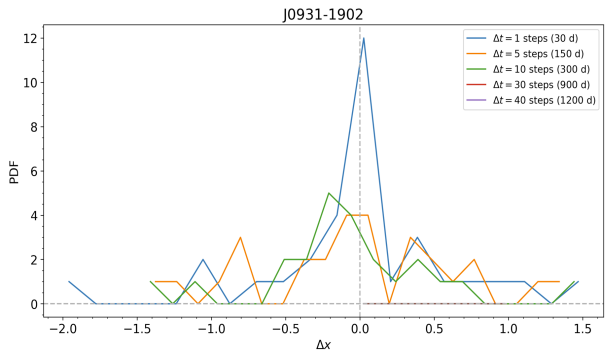
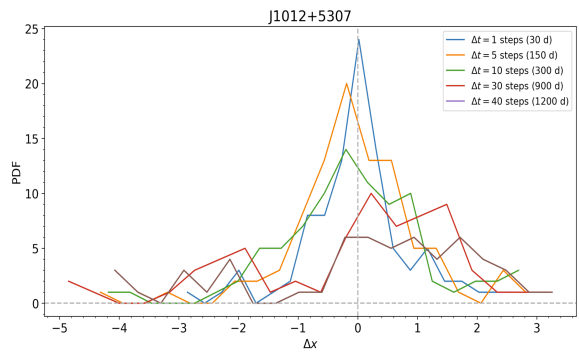
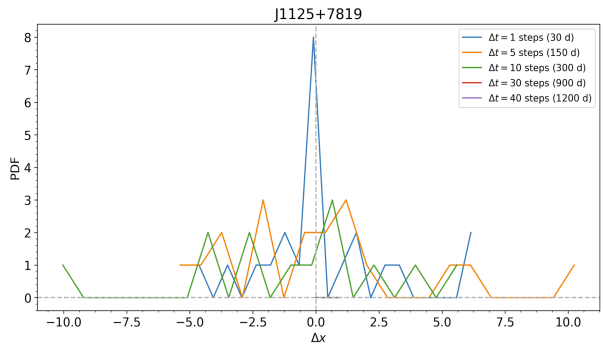
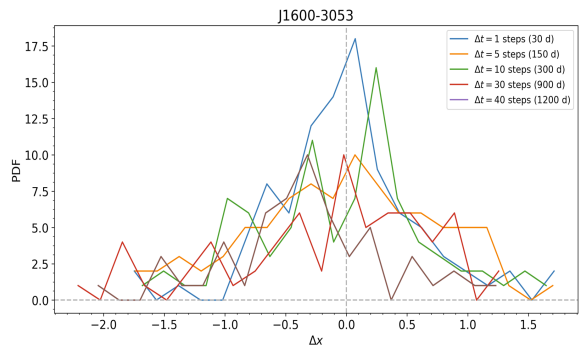
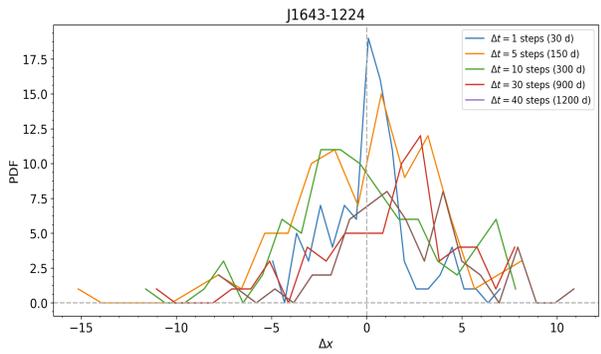
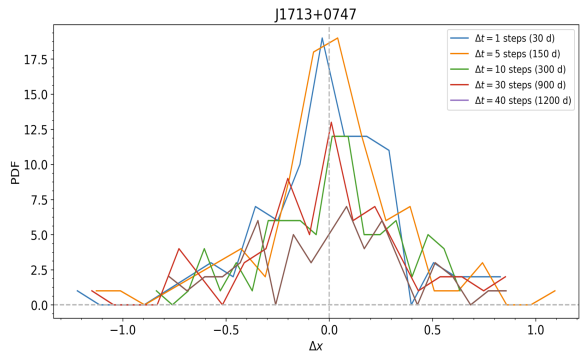
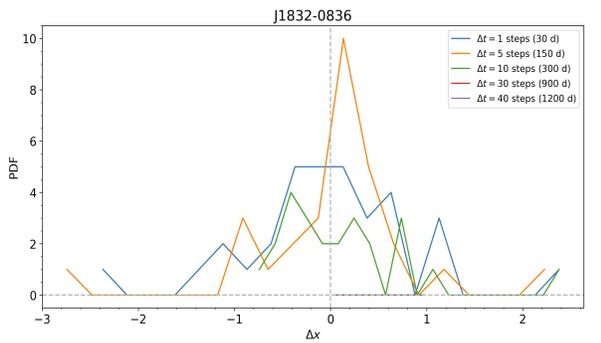
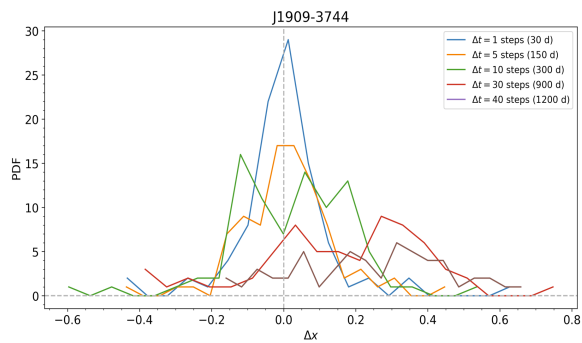

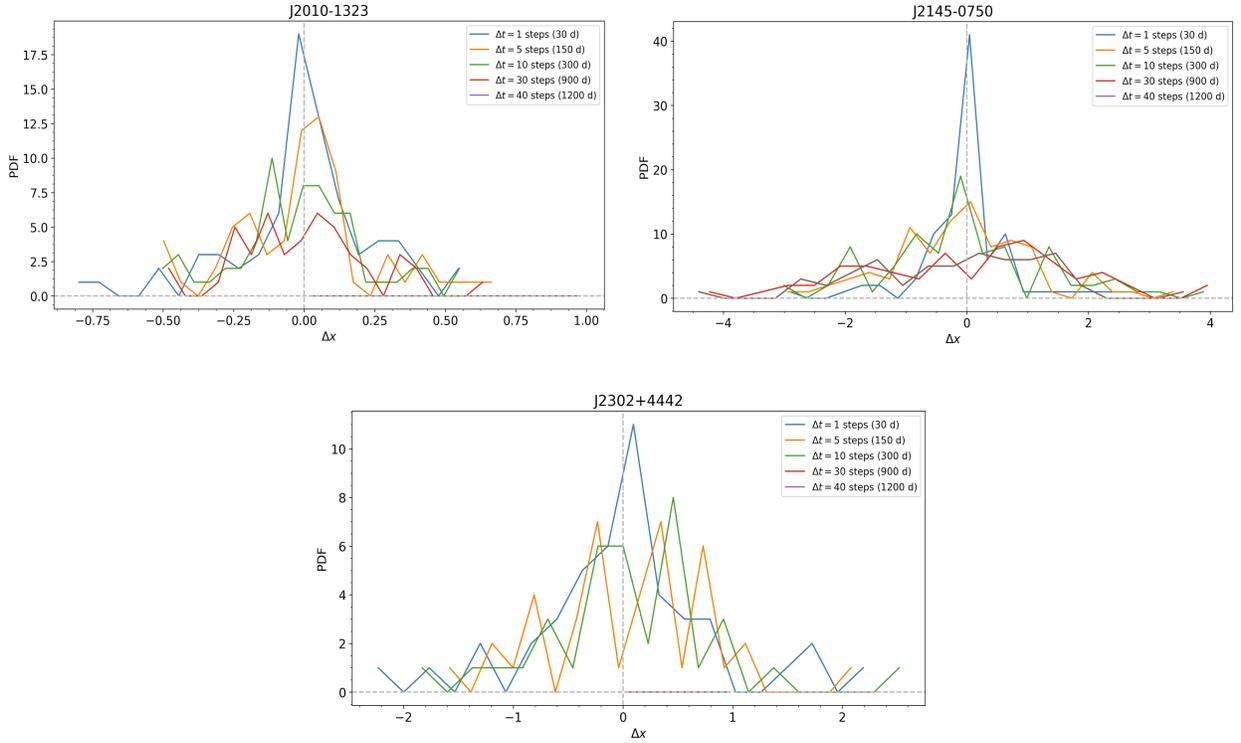

**Figure 5:** Type 2 probability distributions are single-peaked for short lag time and multiple peaked for larger lag times. Pulsars of ths type are: B1937+21, J0340+4130, J0613-0200, J0636+5128, J0931-1902, J1012+5307, J1125+7819, J1600-3053, J1832-0836, J1909-3744, J2010-1323, J2145-0750, and J2302+4442. Of these 15 pulsars, 7 exhibit red noise and 8 do not have red noise.

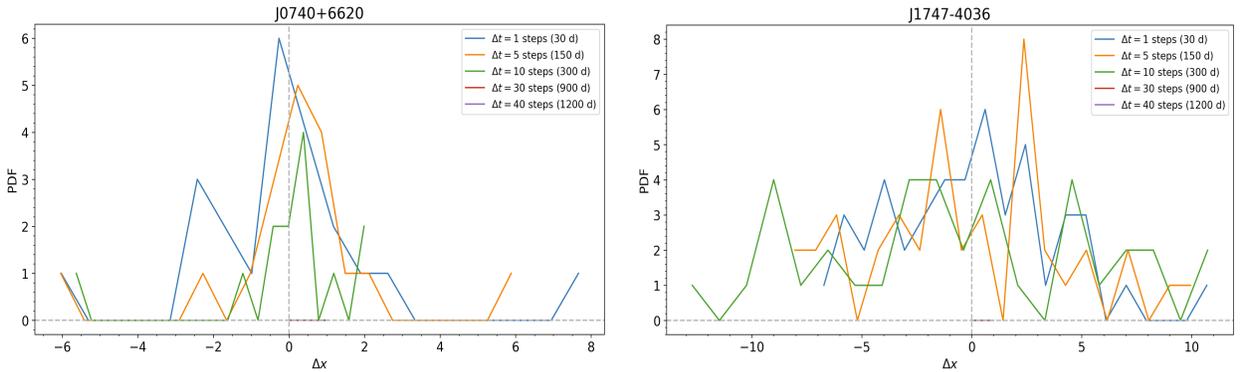

**Figure 6:** Type 3 probability distribution with broad and multi-peaked PDF regardless of lag time. Pulsars of this type are: J0740+6620 and J1747-4036. The former does not have red noise, while the latter does.

### 3.3 Pulsars with Red Noise

We summarize the classifications of pulsars based on their MSD and probability distribution profiles in Table 1. To determine which pulsars exhibit a red noise, we use the work of Arzoumanian et al, (2018). Curiously, the appearance of red noise in pulsars is confined to all those with an increasing MSD.

Of the pulsars investigated, there are also curious features emerging from the PDF. Type 1 PDF always shows a single peak despite the increase in lag times. All pulsars with Type 1 PDF also have a Flat MSD. A Type 1 PDF never exhibits red noise. There are two Type 1 pulsars, however, J1744-1134 and J1918-0642 (marked by asterisks in Table 1), which manifest a double peak at lag time equal to 40 steps, but still broadly peaked around zero. A Type 2 PDF is likewise interesting. It is single-peaked for a short lag time, but develops multiple peaks as the lag time increases. Type 2 PDF may or may not have red noise. An increasing MSD appears to be a required feature for pulsars with red noise. There are two pulsars which do not belong to the above two Types. We label pulsars J0740+6620 and J1747-4036 as Type 3 which is characterized by a broad and multi-peaked PDF regardless of lag time. Table 1 gives a summary where a clear manifestation of red noise is observed for those with an increasing MSD only.

**Table 1**
Profile of PDF and MSD for pulsars with red noise.
The red dots indicate the presence of red noise.

| PDF | FLAT MSD | INCREASING MSD | RED NOISE |
|---|---|---|---|
| Type 1 | J0645+5158 | | |
| Type 1 | J1024-0719 | | |
| Type 1 | J1455-3330 | | |
| Type 1 | J1614-2230 | | |
| Type 1 | J1744-1134* | | |
| Type 1 | J1918-0642* | | |
| Type 2 | J0340+4130 | | |
| Type 2 | J0636+5128 | | |
| Type 2 | J0931-1902 | | |
| Type 2 | J1125+7819 | | |
| Type 2 | J1600-3053 | | |
| Type 2 | J1832-0836 | | |
| Type 2 | J2010-1323 | | |
| Type 2 | J2302+4442 | | |
| Type 2 | | B1937+21 |  |
| Type 2 | | J0613-0200 |  |
| Type 2 | | J1012+5307 |  |
| Type 2 | | J1909-3744 |  |
| Type 2 | | J2145-0750 |  |
| Type 2 | | J1643-1224† |  |
| Type 2 | | J1713+0747‡ |  |

| Type 3 | J1747-4036 |  |
| Type 3 | J0740+6620 | |

## 4 Conclusion

The inherent errors present in timing residuals make it difficult to analyze noise in pulsar time of arrivals. Despite this, we have shown that the classification of timing residuals based on their MSD and probability distribution profiles reveals an underlying pattern that distinguishes pulsars with red noise. In particular, pulsars with an increasing MSD when plotted against lag time all had red noise processes. The results are compelling and demonstrates a rather accurate and straightforward procedure in determining the presence of red noise in pulsars (see, Table 1). This model-free approach can complement existing protocols in the determination of red noise in pulsars which rely on model-dependent statistical procedures with fit parameters (Arzoumanian et al 2018). The unique appearance of red noise for timing residuals with an increasing MSD could provide insights on the nature of neutron stars whose inner dynamics has escaped understanding (Cromartie et al 2020). The continuing improvement of pulsar noise models may also be aided by this distinct MSD behavior observed in timing residuals.

**Conflict of Interest**

*The authors declare that the research was conducted in the absence of any commercial or financial relationships that could be construed as a potential conflict of interest.*

**Author Contributions**

R. R. conceptualized, curated the data, generated all the plots, and reviewed the paper. C. C. B. conceptualized, wrote, and reviewed the paper.

**Acknowledgment**

R. R. wishes to acknowledge the Research Center for Theoretical Physics, CVIF, and Jagna for its natural beauty and rich scientific history that directly inspired this work.

**Data Availability Statement**